\documentclass[11pt,reqno]{amsart}
\usepackage[foot]{amsaddr}
\usepackage[top=1in, bottom=1in, left=1.25in, right=1.25in]{geometry}
\usepackage{graphicx} 
\usepackage{xcolor}
\usepackage[colorlinks=true,allcolors=blue,allbordercolors=white]{hyperref} 
\usepackage{dirtytalk}
\usepackage{fontawesome5}
\usepackage[backend=biber, sorting=none]{biblatex}
\usepackage{pdfpages}
\title[A Quantum of Hope]{A Quantum of Hope: \\ Putting quantum science and technology on stage \faTheaterMasks}
\author{Shawn Skelton$^*$}
\address{$^*$Institute for Theoretical Physics, University of Leibniz Hannover \texttt{shawn.skelton@itp.uni-hannover.de} 
}

\author{Anna Knörr$^\dagger$} 
\address{$^\dagger$Quantum Science \& Engineering, Harvard University \texttt{aknorr@g.harvard.edu} 
}

\author{Jaime-Redondo-Yuste$^\ddagger$}
\address{$^\ddagger$Center of Gravity, Niels Bohr Institute 
}

\author{Manu Srivastava$^\mathsection$}
\address{$^\mathsection$Center for Theoretical Physics -- a Leinweber Institute, Massachusetts Institute of Technology
}

\author{Anna Brandenberger$^\circ$}
\address{$^\circ$Department of Mathematics, Massachusetts Institute of Technology
}

\begin{document}

\begin{abstract}
Public outreach in quantum science and technologies has many goals, ranging from generating interest and dampening hype to making the fascinating and complex topic more accessible. In this work, we present a play on quantum science and technologies aimed at a science-curious audience, which aims to lift the curtains on different researchers' perspectives on this rapidly evolving field. These notes expand on the ideas presented in the play and provide additional sources that may be useful for educators and directors in future productions.

In \textit{A Quantum of Hope}, the end of the world has never looked so bureaucratic. When an alien civilization announces that Earth will be demolished to make way for a research station, humanity is given five years to prove its value. The evaluation criteria? Randomly chosen to be quantum science and technologies. Five experts, sampled from quantum academia and industry, are locked in a room to select a single project proposal to unite global efforts. Douglas Adams meets 12 Angry Men in a galactic-grade grant review committee, where scientific ideas, moral convictions, and unruly egos clash. 
\end{abstract}
\maketitle

\vspace{-\baselineskip}
\begin{figure}[hbtp]
    \centering
    \includegraphics[width=0.45\linewidth]{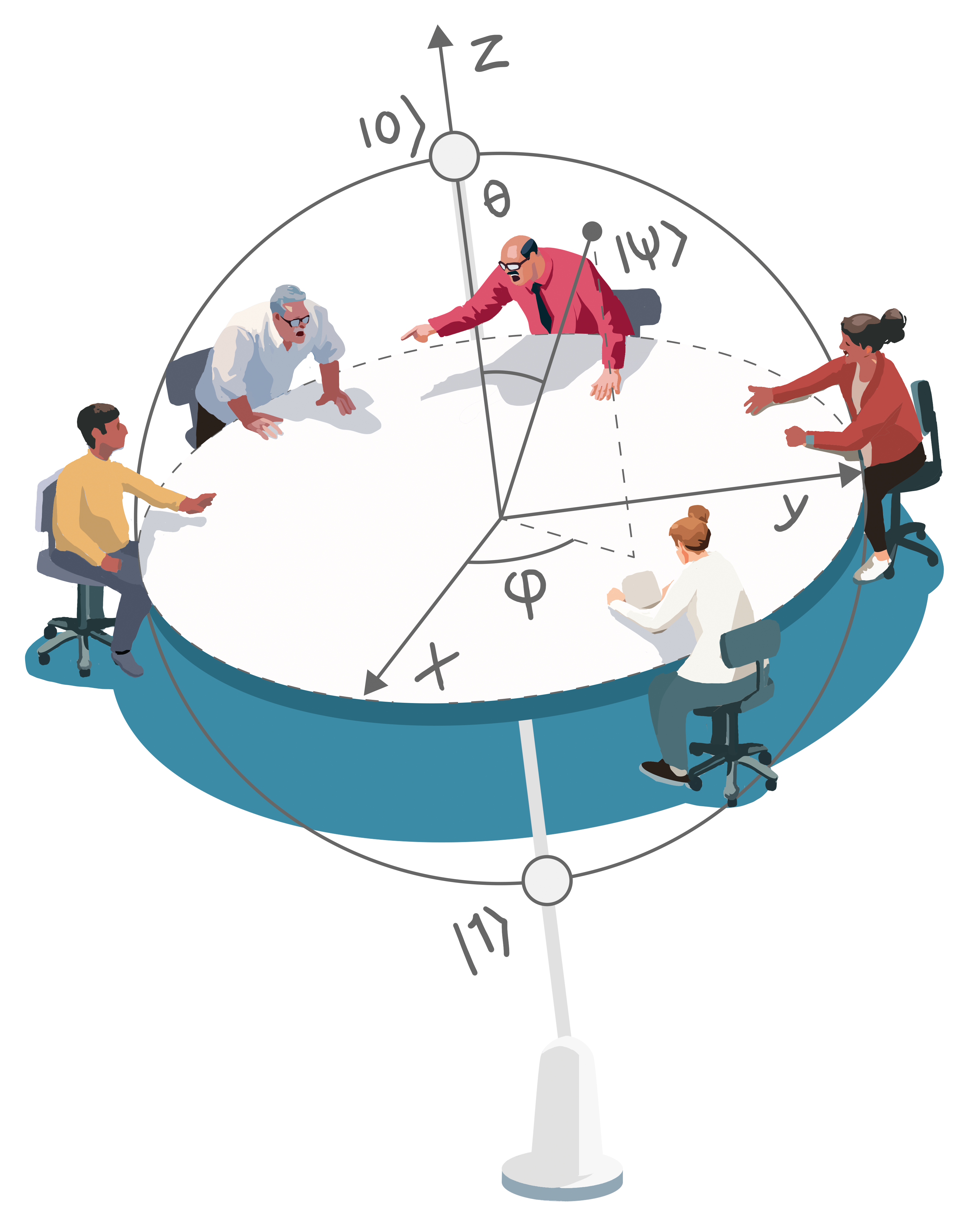}
    \label{fig:bloch-icon}
\end{figure}
\newpage

\section{About the Play}
\textit{A Quantum of Hope} began in June 2024 with the aspiration to write a short story inspired by contemporary physics. After a weekend of plot and character development at our \href{https://perimeterinstitute.ca/news/perimeter-scholars-international-celebrates-its-15th-anniversary}{master's program reunion} at the Perimeter Institute for Theoretical Physics, Canada, we realized that we had pivoted to writing a play. Over the next year, we planned, drafted, and revised \textit{A Quantum of Hope} into its current form. Plot-wise, our script draws inspiration from the film \textit{12 Angry Men} and the absurdity of Douglas Adams' fiction to provide an introduction to the landscape of quantum science and technologies. The play involves many high-level value judgments about the state of this rapidly developing and highly interdisciplinary field, and we tried to capture a range of viewpoints. In this process, we greatly benefited from discussions with colleagues from both quantum and theater backgrounds.

Central to the play are five research proposals, which schematize the questions scientists ask and the objectives technologists hope to accomplish with quantum science. These proposals and the characters' discussions around them are intended as a snapshot of the hopes and doubts, the excitement and uncertainties surrounding quantum research in 2025. Along with the science, our plot also focuses on scientists as human beings, and touches on the intersections of science and technology with wider societal debates. We hope that future readers will find both amusement and value in looking back at these discussions in years to come.

\textit{Disclaimer: Some parts of the ensuing text contain spoilers. Feel free to skip ahead to \hyperlink{PlayPdf}{the play}.}

\section{Themes}
Our play intends to communicate a few key messages about the scientific process as well as the field of quantum science and technologies.

\subsection*{The breadth of quantum research} 
 Quantum-related topics are increasingly present in public media, with quantum computing receiving by far the most attention. We wanted to highlight the broader range of topics that researchers are interested in, by having our characters discuss five different research proposals\footnote{For the sake of the plot, these proposals are presented as distinct research directions, while in reality there is significant overlap and cross-pollination between them. To make up for these somewhat artificial buckets, we do have one character comment on this point towards the end of the play.}. These include noisy intermediate-scale quantum (NISQ) devices and software, quantum foundations, quantum simulations, large-scale universal error-corrected quantum computers, and quantum sensors. Particular attention was paid to distinguishing NISQ devices from fault-tolerant quantum computing (FTQC)  - which in popular discourse sometimes gets subsumed under the umbrella term of \textit{quantum computers}, although critical milestones are still needed to achieve the latter. Many other important topics, notably quantum communication, could have been included. We hope that other researchers with literary interests will take it upon themselves to build stories around these omissions.

\subsection*{The variety of motivations behind quantum research} 
As graduate students immersed in the quantum research community, we wanted to demonstrate the breadth of motivations among quantum researchers. Our characters are somewhat caricatured figures, from the purists who seek understanding of scientific phenomena above all, to those who care, first and foremost, about turning scientific insights into useful technologies. In this field, academic and industry researchers interact significantly, so our characters represent both backgrounds. Managing an ensemble cast for a short play requires a small number of characters, and was more easily done when all characters share a similar technical background. This unfortunately meant excluding non-technical experts, including legal experts, ethicists, government representatives, etc., all of whom play important roles in shaping ecosystems for quantum research to happen within the real world.  

\subsection*{The uncertainty of research} 
\textit{How do you know whether a research program is going to work? On which timescales? How do you choose what research to pursue given this uncertainty?} There is no universally right answer to these questions, and our characters reflect different perspectives one might have on them. All the proposals have a range of merits and challenges, which the characters weigh depending on their motivations and interests. These judgments are at the core of what moves our plot forward, since the characters have to discuss and convince each other which proposals to discard.

This highlights a reality of science that is underrepresented in science communication and outreach \textemdash research is highly uncertain. The results of a research project often look very different from the initial goals of a project, as the team learns from dead ends and unexpectedly productive directions. Hence, when deciding to fund a project, one cannot know for certain whether it will work out (in the manner intended) or not. This is a delicate balance that our play tries to highlight.

Finally, the proposals in this play are constrained by a five-year deadline. Five years to create massive progress in a scientific discipline may sound like an intentionally dramatic choice, and it is. However, the vast majority of academic work operates through funding grants allotted on a three or five-year basis; companies generally publish road maps of their projected progress roughly every five years. Choosing such a tight timeline coupled with the uncertainties outlined above pressure-tests our characters' confidence and raises the dramatic stakes of the play.

\subsection*{The ethics of quantum technologies} 
Given the potential for quantum science to mature into technologies that will impact people's lives, we felt that it was important to include societal questions in the characters' debates. \textit{Who will own powerful quantum computers? At what stage of development do we want to regulate their usage? How does quantum sensing intersect with military applications, and what stance might one take towards these?} Indeed, these questions are already being debated by technical researchers, social scientists, policymakers, and more actors in the real world~\cite{g7statement_2025}, and our characters represent different viewpoints on these issues. Moreover, we want to encourage the interested public to start asking questions about these new technologies. As quantum 2.0\footnote{Quantum 2.0, or \textit{second wave quantum technologies} refers to the current generation of quantum technologies under development, in contrast to earlier technologies such as transistors, which are based on semi-classical physics.} moves out of the lab and into society, how will it be used, and whose vision is being followed? 

These are not questions that can be quickly answered; rather, our play is intended to plant a seed that can grow alongside the maturing technologies. We hope to convey this message by letting the audience vote on the final decision at the end of the play. (We warned you about spoilers!)

\subsection*{Functional understanding of quantum technologies} 
Research in ethics of quantum technologies~\cite{dejong2025functionalunderstandingquantumtechnology} argues that the type of \textit{understanding} that is crucial for being able to participate in debates about the impact of quantum technologies is \textit{not} technical understanding. Rather, one needs to comprehend their \textit{functional capacities}, i.e., what these technologies can do and hence what their potential applications might be\footnote{One might counter that applications can also be quite technical. Yet, in even those cases, one does not need to understand all the underlying physical principles to grasp that these can be harnessed for a specific purpose.}. In alignment with this argument, our play does not focus on explaining the nitty-gritties of quantum mechanics\footnote{If you are interested in shows that explain quantum mechanics, please check out \href{https://www.theatrerenard.com/rebellion-du-minuscule}{La Rebéllion du Miniscule}, written by Antonia Leney-Granger and performed by Théatre du Renard (Montréal)!} and instead dwells on technologies at a higher, more functional level\footnote{Feedback from the theater community also assured us that the audience does not need to understand occasional technical terms (e.g. \say{boson sampling}) used in dialogue, as long as the gist of the argument or joke is clear.}.

\section{The science: some background information}
In the following, we summarize the main points we hope to convey about the five proposals discussed in the play. Although we do not reproduce the full text of each proposal contained in the script, the sentiment of the proposals is summarized in a slogan each\textemdash sometimes with a contentious edge.

\subsection*{NISQ devices}

\textit{`If we make NISQ devices easy to use, the applications will emerge… let's unlock the potential of NISQ together!'}\smallskip

The first proposal read by the characters focuses on writing software for NISQ devices~\cite{Preskill_2018}, in order to use these early-stage prototypes of quantum computers for real-world applications, such as logistics or finance. This proposal is intended to echo an overly optimistic sentiment that has been widely promoted in public media outlets over the past few years, and has ultimately been strongly criticized by the scientific community~\cite{sarmahype}. Concretely, the criticism centers on the hyped promise of soon obtaining a quantum advantage for real-world problems\footnote{Some quantum computers are used commercially today, but computationally, NISQ computers can be easily supplanted by other methods. For one specific example of comparing quantum and classical methods, see \cite{quantumethicsprojectIBMEagleEyed}.} (either a better solution or an equally good but faster solution) by running algorithms on a combination of NISQ computers and classical computers. 

NISQ devices are undoubtedly scientifically interesting, both as a stepping stone towards engineering next-generation quantum computers, for fueling technical debates about defining quantum computational advantages~\cite{huang2025vastworldquantumadvantage, aaronson2025_futurequantumcomputing} as well as for scientific insights into fields such as quantum dynamics \cite{Preskill_2018, PRXQ_SecondQRevolution_2020}. However, using these devices at the current stage for complex problems that industries or governments care about for their day-to-day operations, and obtaining a commercial advantage through some creative software improvements, is an empty hope. That is what our characters quickly convince each other of and what we hope the take-home message for the audience is. Furthermore, the research community is starting to move beyond NISQ devices~\cite{Preskill_2025}, into regimes of quantum computing explored in ensuing proposals.

\subsection*{Quantum foundations}
\textit{`Let's get to the heart of quantum weirdness!'}\smallskip

In over a hundred years that we have been exploring quantum science, it is still controversial how or whether to accept quantum states as part of the "real factual situation" of a system or not \cite{einsteinquantumstate}. More simply, are physicists satisfied with their understanding of how to interpret the mathematics they use every day? What does it mean for measurements to affect the state of quantum objects? Is it satisfactory to have an intrinsically probabilistic understanding of a phenomenon as opposed to a deterministic description?

There are many reasons why one might care about such foundational questions. Plain curiosity\footnote{For example, Anton Zeilinger embraces this view: \say{I can tell you very honestly and proudly, this is good for nothing. I do it for curiosity. It helps to change our view of the world.}~\cite{zeilinger_podcast}}, empirical support for one's preferred philosophy of science, or solving existing open questions in quantum theory. There is also some hope that quantum technologies can provide experimental validation to ideas in quantum foundations~\cite{quantamagazineQuantumParadox,Nurgalieva_2020, Frauchiger_2018, supicselftestingreview2020, Rozema_2024}. Another driving motivation is the difficulties in reconciling our understanding of gravitation with quantum physics. There is hope that research in the foundations of quantum theory may hint towards a resolution to this puzzle.

Our play tiptoes around the fact that researchers and funding organizations sometimes disagree about why foundations research is worth doing. Foundations research is almost never directed towards some well-defined, socially identifiable goal besides pure knowledge or the satisfaction of researchers. Science funders (and many philosophers of science) often counter that the personal feelings of physicists should not really be relevant. The point of funding any highly uncertain or exploratory research agenda, whether it be deep sea explorations or quantum foundations, is usually to spend a small\footnote{"small" with respect to overall science funding.} amount of money on high risk, high reward\footnote{How and whether foundations research counts as "potentially high reward" is really controversial, and usually connects to whether one accepts that physical reductionism (see for example \cite{HalvorsonForthcoming-HALFPA-2}) holds up in quantum science.} gambles. 

Essentially, the play aims to highlight that some models of funding work really well or really badly for different types of science. We personally feel that foundations research does not fit the short-term, centralized premise set up in the play, and so our characters put this proposal aside and move on with the plot.

\subsection*{Quantum simulators}
\textit{`Could simulating superconductivity be the new sexy?' }\smallskip

The third proposal considered by the characters centers on building quantum simulators and using these to solve specific problems in physics, chemistry, and other sciences. This is a narrower but arguably more grounded vision than that of revolutionizing industries using NISQ devices or soon achieving large-scale FTQC, a proposal that follows later. 

In fact, the idea of using quantum machines to simulate complex quantum systems traces back to one of the earliest motivations for quantum computing itself~\cite{feynman1981}, and has seen serious theoretical and experimental progress in recent years~\cite{natureGoalsOpportunities,Altman_2021}. Examples of scientific problems simulators could be used to tackle include dynamics of quantum many-body systems, probing high-temperature superconductors, and solving lattice gauge theories \cite{Meth:2023wzd,Banuls:2019bmf,Funcke:2023jbq}. 

However, insights from quantum simulators could still impact both our understanding of the nature of the world, as well as potentially translate into advances in fields like material science further down the road. We also highlight the epistemic novelty of quantum simulators. They can function either as calculators or experiments, giving insights into similar but more complicated quantum systems. For an in-depth discussion of quantum simulators as a new instrument for scientific understanding, please refer to~\cite{Hangleiter_2022, johnson2014_whatisqsimulator}.

While still facing significant experimental challenges, the path to a practically useful quantum simulator is considered more feasible than that to fully FTQC~\cite{DALEYpractical2022, trivedi2023quantumadvantagestabilityerrors}. However, quantum simulators lack the media sparkle of full-blown quantum computational power and imminent commercial disruption, and have thus received far less public attention. 

\subsection*{Large-scale universal error-corrected quantum computers}

\textit{`The hurdles to quantum computing are engineering challenges, and can be overcome.' }\smallskip

The fourth proposal on the table focuses on the holy grail goal of building a fault-tolerant quantum computer, a device capable of solving a wide range of meaningful problems in science and industry. Unlike the NISQ proposal, which emphasizes creative software solutions for noisy prototypes, this one zeroes in on the hardware challenge, that is, how to construct systems with extremely low error rates~\cite{riverlaneQuantumComputers}. It is worth emphasizing that achieving the feat of engineering error correction on large-scale systems is arguably one of the most formidable scientific and technical undertakings of our time.

Leading academic institutions and big tech companies alike are investing heavily in this direction, and in the past few years, some remarkable breakthroughs have pushed the field forward~\cite{physicsworld2024_qec}. Still, only the most optimistic of our characters is confident about the idea of seeing a large-scale fault-tolerant device within a five-year window. In order to make the distinction between full-blown quantum computers and quantum simulators clear, our characters dwell on the trade-off between the remarkable hardware challenges of achieving FTQC, contrasted against the more limited application range of simulators\footnote{Of course, universal quantum computers could also be used for quantum simulations, but splitting the proposals up in this way allowed us to contrast the engineering hurdles and areas of application more clearly.}.

Further, this proposal prompts a deeper conversation in the play, not just about feasibility, but about governance \cite{g7statement_2025}. \textit{Who will own such machines? What happens when one state or corporation has access to them before others? What is the relationship between industry and academia in this field? What kind of security, economic, and ethical frameworks do we need to develop now, before the hardware actually arrives?} While the characters pose each other these questions, answering them would require a play in itself (or rather more!).\footnote{This point requires some delicacy because \say{quantum ethics} superficially sounds as hyped as quantum computing. Technical experts often rightfully balk at the notion of speculative governance because they worry about creating hype or high expectations when trying to anticipate the effects of non-existent technologies. However, non-technical actors are rightfully concerned that, unless integrating technology ethics is done within early design phases, it will become severely challenging to regulate for the first decades of its use, both in cases where quantum technologies might be grandfathered into existing regulations as well as when new regulations are needed. Taken together, these concerns are also known as the Collingridge dilemma~\cite{collingridge1982_socialcontrol}. We do not claim to have precisely the correct balance between these concerns, but strongly suspect it requires earnest and in-depth discussion between technical, industry, and governmental experts.} 

\subsection*{Quantum sensors}
\textit{`By uplifting sensors and scientific measurements, we uplift all science!'}

A quantum sensor is any one of many potential types of scientific measurement devices that rely on quantum mechanical properties to work. Since \say{quantum sensor} is a broad umbrella term, some already exist today. One may argue that quantum sensors have been around for decades, e.g., in atomic clocks \cite{atomicclock}, or medical instruments, including MRI technology~\cite{degenqsensing2017}. However, quantum sensors that use the quantum mechanical property of \textit{entanglement} are an active and currently developing research topic. There are many proposed quantum sensors, several of which use designs similar to those that are being explored for quantum computing. Such gadgets could one day provide better measurements of many different quantities: magnetic, electric, or gravitational fields, as well as time, pressure, and temperature. 

Some sensors are being explicitly developed for commercial applications with the aspiration that they can work at room temperatures outside laboratory settings. Advanced quantum sensors could be used in biomedical research and clinical practice  \cite{natureQuantumSensors, TAYLOR20161}, mining prospects, and navigation~\cite{ohperspectivesensors2024}. For a less technical overview of different types of quantum sensors, see the introductory sections of the \say{quantum sensing} chapter of ~\cite{ezratty2024understandingquantumtechnologies2024}.

Importantly, quantum sensors are already contributing to other areas of science by increasing measurement precision, notably by being integrated into gravitational wave detectors \cite{caltechLIGOSurpasses,PhysRevX.13.041021}. This brings forth the discussion on whether progress in science is driven by theoretical developments which propose novel experiments, or by improvements in precision measurements\footnote{For example, Michelson famously claimed about this debate that "future truths of physical science are to be looked for in the sixth place of decimals''~\cite{UChicagoAnnualRegister1897}.} which may hint towards theoretical inconsistencies in our models.

\section{A note on voting}
In case you are interested in producing our play, we'd like to draw your attention to (spoiler alert!) the audience voting that occurs at the end of the play. Our website contains a section where we are compiling the voting outcomes of ideally all productions. You could think of this as collecting the measurement outcomes of the audience wavefunction \faSmileWink[regular]. Do let us know what the results of your production were!

\section*{Acknowledgments}
We are grateful to many friends and colleagues for reading early versions of this work. Among the many readers, a few in particular contributed expert guidance on specific topics we covered. For the quantum foundations section, we are particularity grateful to Henrik Wilming and Martin Pl{\'a}vala; for the quantum simulators section, Dominik Hangleiter; for the quantum sensors section, Ben MacLellan. We are grateful to Susanne Yelin, Evelyn Hu, Rodrigo Araiza-Bravo, and Olivier Ezratty for general scientific feedback, and to Kelly Werker Smith for feedback on the preamble.

This project was supported by generous funding from the Perimeter Institute for Theoretical Physics and Two Small Fish Ventures. We thank Sunny Tsang at Perimeter Institute for facilitating the collaboration. 

Amanda Clarke did a brilliant job with structural and copy edits of the script. \href{https://www.innovailia.com/}{Innovailia} and \href{https://www.bramastudios.com/}{Nic Brama} facilitated our production poster designs.
 
Finally, we thank Zivy Hardy and Adrienne Dandy for organizing table work sessions for the play's first draft, as well as all of the Perimeter residents who participated in these sessions and provided feedback. 

\vfill 
\centering
And now... enjoy reading the \hyperlink{PlayPdf}{play}! 
\vfill 

\printbibliography

\vfill
\hypertarget{PlayPdf}{}


\section*{Supplementary material}

A QUANTUM OF HOPE

---

A Play in One Act

by

Shawn E. S. Skelton,
Anna Knörr,
Jaime Redondo-Yuste,
Manu Srivastava, and
Anna Brandenberger

*written during the International Year of Quantum Science & Technology 2025*

info@aquantumofhope.com
www.aquantumofhope.com

In memory of Ray Laflamme - who cast rays of insight into quantum theory and rays of humor, guidance, and kindness into the lives of many in Waterloo and beyond.

CAST OF CHARACTERS

| | |
|---|---|
| PUCK | A helpful fairy residing on Earth with a vast network of human friends and the power to suspend disbelief. At least 500. |
| MIGUEL | A young and soft-spoken professor. As insecure as he is passionate. Early 30s. |
| PROF. SCHMIDT | An established professor of theoretical physics, driven by curiosity. 65+. |
| ELISABETH | A mid-career professor with a grouchy, no-nonsense persona and a strong moral compass. Late 40s. |
| JEFF | A seasoned startup founder, behind on the latest research outside of his field, but optimistic and eager to contribute. Late 30s. |
| STEF | An energetic industry leader on a mission, with little regard for other's opinions. Late 40s. |

SCENE
A meeting room. On one side, a blackboard partially filled with
gibberish. In the center, a table and chairs that can be moved
around easily. Posters of the different proposals are hanging on
the walls. Five folders (only one of them open) and an envelope
lay on the table.


                                TIME
A single day.


                             SINGLE ACT
SCENE 1: The news arrives
SCENE 2: At the meeting room
SCENE 3: NISQ devices
SCENE 4: Quantum foundations
SCENE 5: First coffee break
SCENE 6: Quantum simulators
SCENE 7: Quantum computers
SCENE 8: Second coffee break
SCENE 9: Quantum sensors
SCENE 10: The verdict

## Scene 1: The news arrives

SETTING:              An empty stage, lit only by a spotlight.

AT RISE:              Puck steps into the center of the spotlight, waiting impatiently, nervous but trying to put up a confident façade. Puck notices the audience, walks towards them, and begins.

                          PUCK
It all started with light. Matarisvan stole the hidden fire from the gods and gave it to the humans. Or Prometheus or Maui or any one of the other fire-stealing champions. For thousands of years, fire was protection. Fire was warmth and fire was power. It was also a place for music and storytelling. Humans gathered around it and turned its light into art that became language that morphed into stories which revealed new secrets of the world. They asked why and how and what if? And humans, they are relentless. They figured out the why and the how and the what of many things until fire burned inside every house; in every street lamp, a tiny fire shrouding the night sky and what lay beyond from the eyes of this curious, relentless flock. But humans were not meant to remain alone forever. Another flock from beyond the stars has taken notice. (more visibly nervous) Now what do they want?

   (A masked figure, dressed in black, hands an envelope to Puck.
         They skim it, speaking some passages aloud.)

Environmental impact assessment notice… Construction of an observatory with high scientific value… Subject to a determination of minimal ecosystem impact, hereinafter defined, the planet locally known as "Earth" and its associated satellite, known as "Moon," shall be slated for demolition…

               (Pause to continue reading.)

Must show evidence of an equivalent cultural or scientific value… Evaluation criteria have been selected randomly to minimize cultural preference bias…

                    (with alarm)

Here chosen to be quantum science and technologies… Evaluation to be completed five years from date of receipt.
> (to the audience)

They're joking. They better be joking… But they really don't joke…

> (Looks down to read more of the letter, then turns to the audience, speaking in the manner of one trying to deliver bad news with a bit of positivity.)

Be at good ease still. Usually such matters can be resolved with the right strategic chat with the being in charge. Just as long as I get there before any kind of public announcement.

> (Theatrical sirens blaring, or other stagecraft, to indicate that a communication has gone out to the rest of Earth. Puck starts pacing around the stage, visibly angry.)

Why did they not reach out to me? Robin Goodfellow, prince of the fairy court, sprite of the planet Earth? But no, they had to broadcast their message everywhere… The entire planet will overreact, and it will be twice as difficult to untangle this mess!

> (Calms down, theatrically discards the letter.)

I'll grant them one thing – I've always liked a challenge of wits. It should have been easy to prove the value of the culture produced on this world. But yet again these creatures disappoint me. Of all things – blues music, satirical literature, free-verse poetry, mediocre legal TV series – all the things that are shared across the galaxy and beyond, they choose QUANTUM!

> (shakes their head)

I could have spoken of artistic triumphs or deep-sea exploration or any of a number of things. But very well. What is Quantum? We were talking about the fire, right?
As humans chased after stories, they eventually discovered that fire did not give light like the unbroken flow of a waterfall into a pond but like drops of water falling, one by one, onto one's face on a rainy day. Always individual in essence, even when collective in effect. Quickly humans identified each of

these building blocks, called photons, each of them a raindrop, which their devices could detect. Humans invented new math to talk about this phenomenon, and they baptized it **quantum**. They peered inside the building blocks themselves, peeking into the primordial fire. They learned to harness the power of quantum and to interpret how raindrops collect into threads of water. And so they continued, chasing after the smallest raindrop, peering deeper and deeper in their relentless effort to understand. Tonight, dear spectators, you will witness firsthand the frontier of these spectacular developments. Together we shall behold what the future may hold for quantum science. What technologies will we be promised? And will they be enough to avert Earth's fate?

END OF SCENE

<u>Scene 2: At the meeting room</u>

SETTING:                At the meeting room

AT RISE:                Schmidt is alone on stage, skimming
                        through the dossier of proposals, and
                        scribbling gibberish on the blackboard,
                        concentrating. Miguel approaches him.

                         MIGUEL
Prof. Schmidt?

    (Schmidt does not react; Miguel clears his throat.)

Prof. Dr. Schmidt?

        (Schmidt turns; they shake hands.)

I'm Miguel. We met some time ago in Geneva, although you
probably don't remember. (chuckles nervously) It's a pleasure to
see you here again!

                         SCHMIDT
              (with surprise)
Oh, okay. (looks up, trying to recall) Ah, Miguel! The newest
faculty at Cambridge. Of course I remember you and that
wonderful seminar you gave. Spin-motion coupling in optical
traps, right? All good with you?

                         MIGUEL
              (proudly)
Yes, I was working on that! Before the end of the world, and all
that, you know. I'm doing okay, given the circumstances.

    (Conversation moves to the background as Jeff and Elisabeth
                       enter together.)

                         ELISABETH
              (awkwardly letting Jeff go first)
After you.

                         JEFF
Hi! Lovely to meet you. Elisabeth, right? I read your quantum
key distribution paper. Brilliant work.

                    ELISABETH
Oh yes, everyone seems to have read that one or at least skimmed it. Do you work on quantum communications? Sorry, I can't say that I recognize you.

                    JEFF
Jeff! You wouldn't have heard of me. I don't publish much myself, but perhaps you've heard of my startup? We work on quantum sensors, mostly magnetometers.

                    ELISABETH
Oh, I see. Well, I can't say I follow the industry side very closely. If you'll excuse me, there is a colleague I should catch up with…

   (Elisabeth approaches Schmidt and Miguel, engaging in lively
    conversation. Jeff takes a seat. Stef enters confidently and
                      approaches the table.)

                    STEF
Jeff Dunphries? An honor to finally meet you in person. I've heard great things about you.

                    JEFF
              (nervously)
No, no, Stef, the honor is mine! I've always been a big fan, you know? Opening pathways for physicists in industry — that Ted talk you gave.

                    STEF
Don't even mention that! I was so young!

                    JEFF
But it was great! It actually inspired me to push towards a path outside academia. That was *so* not for me.

                    STEF
So not for most people with talent… Academia…

                    JEFF
And so, here I am! I followed your advice and thrived in industry till I could fund my startup and pursue my own ideas!

(Elisabeth approaches Jeff from behind.)

JEFF
I'd rather not be involved with aliens and the Earth being blown up, but I guess you can't have it all in life.

ELISABETH
(clears her throat)
So, should we get started? There is a letter here on the table from Puck, a friend of mine!

MIGUEL
Wait, Puck? One of my best friends is called Puck? By any chance, are they really into trance music?

ELISABETH
I would know that how?

STEF
There can't be that many Pucks. Does this friend of yours enjoy alpine hiking?

MIGUEL
Wait, are you saying you also…? Do we all know a Puck?

(Nods all around.)

MIGUEL
Charismatic and connected? Always seems to know what you're talking about, no matter the subject?

JEFF
Bakes a mean matcha Madeleine?

SCHMIDT
It just figures that Puck would have gotten themselves into this one. (pause) That explains why we're all here! I was wondering why some of you were from, shall we say, non-traditional backgrounds. (Gives Stef and Jeff a measured look.)

ELISABETH
Shall I read it? Dear friends, how lovely to welcome you all to Geneva! Brought together by this momentous moment in our

planet's history. I'm sure you'll do a wonderful job saving Earth and humanity and the mice and all.

    (Murmuring among the council members. They look at each other uncertainly. Elisabeth glares at them. They quiet down.)

                        ELISABETH

As you all know, aliens are planning to build a galactic science station in the most unfortunate place — right where Earth is located. Graciously, they have given Earth five years to prove that Earthen science is more valuable than this science station of theirs. They appear to be interested in quantum science and technology. We must dumbfound their galactic wits with impressive results, lest Earth be destroyed! And you, my sextet of dear friends, have a crucial role to play in this story—

                        MIGUEL
          (confused)

Wait, sextet? Are we missing a sixth person?

                        SCHMIDT

I did have a colleague from Tsinghua University in China who told me she would be joining.

                        MIGUEL

Should we wait for her?

                        SCHMIDT

No, that'd be of no use. She couldn't get a visa in time.

                        ELISABETH
          (puts the letter down, resigned and cynical)

Some geopolitics transcend global threats.

                        MIGUEL

So, we're down to a quintet.

                        STEF
          (authoritative)

Five people are more than enough.

                        ELISABETH
          (picks up the letter again)

I have handpicked you to form the most important grant review committee in history! Your task is to fish out the most promising quantum research proposal to be completed in the next five years. The winning proposal will be given near-unlimited financial support from nations across the globe.

(Jeff inspects one of the binders.)

ELISABETH
Here are the proposals I have gathered in the last week from leading scientists. Obviously this is of utmost urgency. You must get started as soon as you can. Until you have made a unanimous decision, you shall not leave this room! Good luck. And please, don't get this one wrong!

(A moment of dumbstruck silence.)

JEFF
(trying to appear calm)
Can I just say, it's an honor to be here with you all.

SCHMIDT
(dryly)
I suppose you can.

MIGUEL
It's a great responsibility, certainly.

ELISABETH
Yes, yes, very nice to be on the "end of the world committee" with you all.

JEFF
It's crazy, isn't it?

ELISABETH
What is?

JEFF
Well… Aliens? Visiting Earth? Them being as enthusiastic about quantum as we are?

STEF

I'm not surprised they chose the quantum frontier, but I still prefer quantum without aliens and an Earth-ending threat. It's beyond my comfort threshold, that's for sure.

> ELISABETH
> (snarky)

And we know that threshold is pretty high… Look, let's just get on with this.

> MIGUEL
> (earnestly)

I think it's good to acknowledge our feelings first. Last week I was sitting at my favorite roastery in Cambridge grading the undergrads' quantum mechanics midterms, and now I'm grading the world's most promising dreams in quantum! So, you know, I am scared.

> SCHMIDT

Fear is not going to get us anywhere. Let's get started.

> JEFF

How are we even supposed to judge these? I mean, do the aliens have crazy advanced quantum technology, and just want to see us sweat? Or do they know only the very basics? If they have totally different ways of explaining it, will they even understand what we present? I get confused trying to explain quantum mechanics to my own family, never mind aliens!

> SCHMIDT

Why? It's quite simple. When you have an infinite-dimensional Hilbert space with non-commuting operators, the paradoxical behavior just jumps out! My grandchildren can understand that!

> ELISABETH

Maybe yours can. I usually just say it's the study of light, and the particles that combine to create atoms. And of course, also the study of how light and atoms interact. Which means that it's the theory of how all material properties are possible.

> MIGUEL

And now we know enough to start playing with arrays of individual atoms in a lab, which might have technological applications.

ELISABETH
Which is why we all have jobs.

STEF
(with enough authority to finally get the group started)
Speaking of jobs, we have one to do. Time to select the most ambitious project humanity can dream of. What do we have here… (picks up the top file, her face falls) Unlock the NISQ[1] potential now!

ELISABETH
Okay, so we're all going to die.

JEFF
Hey, come on, if the proposal made it here, we've got a responsibility to consider it.

ELISABETH
(looking towards Stef)
Or somebody had deep pockets to push it this far.

STEF
You can't think I endorse this shortsighted, unambitious, cherry-picking—

JEFF
Wait, wait — can we just have a look together, please?

MIGUEL
(picking up the first file)
Oh look, here's the summary!

END OF SCENE

---

[1] NISQ rhymes with disk.

<u>Scene 3: NISQ devices</u>

(Miguel stands up and starts reading, pacing around the table.)

                        MIGUEL

Introducing the future of computing: **NISQ** – the cutting-edge technology revolutionizing industries around the world! **Noisy Intermediate-Scale Quantum** devices are the quantum computers sitting in front of our noses. Some will say that because they are too noisy, too prone to errors to run most quantum algorithms, they can't solve interesting real-world problems. But this is a failure of imagination. This proposal will create **software and algorithms** with which NISQ devices can begin to pull their weight and unlock quantum speed-ups for problems that businesses care about.

(Miguel reaches Jeff, drops the proposal in front of him, and walks back to his chair.)

                        JEFF
                  (enthusiastically)

Our user-friendly! open-source! software packages will allow end users and domain experts across the world to explore new quantum applications on the quantum computers that exist in the world today. We're ready to supercharge industries like **finance**, **logistics**, and more.

(Elisabeth reaches towards Jeff.)

                      ELISABETH
                  (disgusted)

In **finance**, NISQ devices can redefine risk modeling, asset pricing, and portfolio optimization. With quantum-enhanced calculations, financial institutions can process vast amounts of data in ways that traditional computers simply cannot. (Makes an exasperated gesture and hands the proposal to Stef.)

                      STEF
              (sarcastic tone, laughing)

This means more accurate predictions, smarter investment strategies, and a stronger edge in a competitive market. In other words, we're ready to do business the quantum way, on a galactic scale. We have good enough quantum computers right now;

all we need is the right software and a push for applications. **Let's unlock the NISQ potential together** today! Woo!

(End of proposal.)

SCHMIDT
I had hoped for more promising material than this.

MIGUEL
Apart from the technical nonsense and buzzword soup, who assumed aliens would be impressed by, you know… capitalism?

ELISABETH
If I had a nickel for every hour wasted in a committee meeting when some fool thought they could get a quantum advantage in a week…

SCHMIDT
You'd be almost as rich as Stef!

JEFF
(with hesitation)
Let's be scientific about this. What exactly is so wrong with this proposal? I mean, we don't have perfect quantum computers, and we have only five years, so?

ELISABETH
Classical computers can already do everything that those NISQ computers can do, regardless of whatever fancy software you create for them. Actually, even my laptop can do everything better.

STEF
And cheaper.

SCHMIDT
Except for boson sampling.

JEFF
See! There is at least one thing.

ELISABETH

Oh, fine, my apologies to boson sampling. Let me rephrase. My laptop can do everything *useful* that those NISQ computers can, only better.

                        MIGUEL

Let's be serious for a moment. You don't think there's any chance, Elisabeth? I mean, if we make it easier for anyone to use NISQ devices and put documentation online, you know, make it actually accessible, maybe someone comes up with an algorithm we haven't thought of? Or some way of applying boson sampling in the real world? Some important problems that could be solved even without perfectly working quantum computers.

                      ELISABETH

I think that if you give everyone a bunch of LEGOs, then someone is going to build a three-meter Death Star. But I'm equally sure that no one is going to build a functional International Space Station.

                      SCHMIDT

I'm lost here.

                        MIGUEL

Me too. We are all scientists. Should we not believe in the power of human curiosity and creativity?

                    ELISABETH

I don't disagree with that, but I think the tools are just not good enough. (Leans forwards into the table and launches into the explanation with boredom.) Back to my analogy: to build the ISS you need titanium alloys and flame resistant composite materials. But every step of how you put those materials together is precisely engineered. With NISQ devices, we know what materials are useful, but we haven't quite figured out how to engineer them into a large system with low error rates. The only algorithms we can run on NISQ computers are stunningly simple, and that's just not good enough yet for these applications!

                        MIGUEL

Right… kind of like trying to do space experiments on a wobbly prototype of the ISS that can't even make it to space!

                          JEFF

>                    (moving uneasily in his chair)
> Surely you aren't saying that every major tech company in the
> world is throwing money at the quantum equivalent of LEGO?
>
>                              STEF
>                (leaning forward, speaking to Jeff)
> Certainly not! Everything we learn about NISQ devices paves the
> way for bigger and better devices down the road. (directed at
> Elisabeth) But they're just the journey, not the destination.
>
>                            ELISABETH
> The journey is part of the destination, Stef. (Takes a breath and
> leans back in her chair.) Look, we all probably agree that NISQ
> devices are scientifically interesting.
>
>                             MIGUEL
> Yeah, we learn a lot from controlling and measuring them.
>
>                            ELISABETH
> But making algorithms simple enough to run on them and thinking
> you can apply them to solve finance problems?! I mean, you might
> as well try to gene sequence a dinosaur with quantum technology.
>
>                              STEF
> So we're all in agreement then. Let's vote and be done with it.
> All in favor of rejecting this proposal, raise your hands.
>
>      (Schmidt, Stef, and Elisabeth raise their hands immediately.
>          Jeff takes a moment, but eventually follows suit.)
>
>                             MIGUEL
> Shouldn't the vote be anonymous? You know, to avoid peer
> pressure and all of that?
>
>                            SCHMIDT
>                            (to Jeff)
> Do you feel peer pressured?
>
>                  (Jeff sighs, shakes his head.)
>
>                            SCHMIDT
> Anyone else?
>
>             (Silence. Those with raised hands wait.)

                    MIGUEL
Okay. (raises his hand)

                   END OF SCENE

<u>Scene 4: Quantum foundations</u>

                             SCHMIDT
          (stands and picks the next proposal)
What's next? Ah yes, this is far better!

*(Begins to read, walking around like he is delivering a lecture. Other council members try to take over at points throughout the proposal, but Schmidt is too absorbed to notice.)*

                             SCHMIDT
Go beyond the flashy technology, and you'll find that the beating heart of quantum science is intellectual curiosity and philosophical wonder at the natural world. (visibly excited) One hundred years ago, scientists were fascinated by the physics of light and atoms. This led them to quantum theory, the rules that govern all the elementary particles in our universe, revolutionizing not only science but also philosophy.

                              JEFF
                     (quietly)
Oh my god…

*(Schmidt stares at Jeff for a second, then continues reading the proposal.)*

                             SCHMIDT
It is time to rekindle this aspiration. It is time to push beyond mere applications of quantum theory, to grapple with our assumptions about reality, to reignite the quest to resolve century-standing paradoxes in quantum theory and tackle the unification of quantum mechanics with Einstein's theory of gravitation.
That's what should keep us up at night. Let's bring innovative minds together, in meaningful dialogue, designing and building experiments to test the core of quantum theory itself. John Bell's work showed the way. Let us not worry about the applications of this **quest for truth**. Let us follow the beacon of pure insight through the wilderness of confusion. For the public and for scientists at large, we can put the FUN back in fundamental physics. **So let's get to the heart of quantum weirdness**!

*(End of proposal.)*

                    SCHMIDT
               (emotionally)
I am very encouraged by this proposal.

                    ELISABETH
               (half exasperated, half fond)
Of course you are.

                    SCHMIDT
When I was in grad school we did not have so many visitors. No zoom meetings, nothing like that. We'd have to wait weeks or months to read papers published on the other side of the world. Anyway, don't let this old man rumble. But you should understand. Back then, we had time to think about what got us really excited, the questions that made our eyes spark! A quantum theory of gravity, the problem of measurement, hidden variables… I fear it's not like that anymore, at least not with my students. But those questions… Those are the deep ones!

                    STEF
I think Prof. Schmidt raises a good point. Perhaps if we actually understand all this quantum nonsense, that could impress the aliens. Show that we are "civilized."

                    MIGUEL
What does civilized mean though?

   (Brief moment of confusion as Schmidt finally takes a seat.)

                    JEFF
That's right! How do we know what another species, or for all we know a bunch of different species, will find impressive?

                    ELISABETH
Maybe they prefer science with a clear-cut path to improving life for everyone on the planet. Penicillin — now that's impressive. Or the smallpox vaccine.

                    MIGUEL
Except, you know, the aliens don't seem too concerned about life on Earth.

                    SCHMIDT

They are willing to blow up a whole inhabited planet, but they are also willing to give us a chance. What does that say in terms of their morals?

JEFF
We don't know whether they would ever consider us as their equals…

STEF
Maybe they see us the way we see a fluffy golden retriever.

SCHMIDT
Or a cow on its way to the slaughterhouse.

MIGUEL
On the other hand, they did ask us to prove our understanding of the quantum world. So somehow our understanding of what is fundamental cannot be too different.

ELISABETH
(frustrated)
We're getting distracted! If Puck wanted us to analyze the social dynamics of these aliens, surely they would have chosen someone else. (teasing wryly, to Schmidt) Unless you have picked up a sociology degree since Barcelona?

SCHMIDT
I have not, but it's worth thinking about, isn't it? Aren't you curious, Elisabeth?

MIGUEL
Elisabeth is right. We should stick to the science. That's what we do. Other people can examine the social aspects better than us.

(There is a brief pause while everyone thinks. Jeff rereads the proposal.)

MIGUEL
This brings me to my question: Does anyone here seriously think explaining quantum foundations is as simple as throwing money at the problem?

STEF

You might be surprised at just how much can get done when excellent minds are organized efficiently and gathered under strong leadership. Just look at Google or Microsoft.

                    SCHMIDT
Or the Manhattan project! Now that was a productive environment.

                    STEF
I wouldn't have gone there, but now that you bring it up, sure.

                    MIGUEL
           (concerned)
What I hear you saying is, we have a comparable amount of support, for very very different goals. Right?

                    ELISABETH
           (resigned)
One Manhattan project reference and we haven't even started talking about full blown quantum computers yet… I'm so proud of us.

           (A brief pause.)

                    JEFF
           (diffusing)
Okay, in any case, all those projects had a clear vision and leadership structure. This proposal is… philosophy.

                    SCHMIDT
So what if it is? Quantum foundations is also proper physics, full stop.

                    MIGUEL
I see that. But it takes imagination, too, and we really don't know what strategy will succeed. (pauses)

           (The group looks towards Schmidt.)

                    MIGUEL
The question is, do we really believe that a hundred years of scientific exploration can be resolved with a five-year plan?

                    SCHMIDT
           (shrinking in his chair)

Well, I have to admit it is a little risky. But is that really the problem? Nothing worth doing is without risk.

ELISABETH

Do you want to stake the world on it? Can anyone explain what makes good foundations research? Does anyone know how you find a good idea or how to ensure that someone has them regularly?

MIGUEL

I agree. Like it or not, sometimes science needs creativity, and creativity is stymied by pressure. You don't just sit around and wait for a world changing idea. You work on stuff, you do things that are useful, and maybe an idea comes after a year or after twenty. Or never!

(All eyes are on Schmidt. His face turns red, his breath speeds up.)

SCHMIDT

Then what are we even doing here? Nothing worth doing works on a five-year timeline. Not particle physics, not genetics, not even space exploration! It's ridiculous to talk about doing impressive science without long-term planning and long-standing curiosity. We spend enough of our time chasing after three-year or five-year grants, and if we are going to spend the last five years on this planet doing research, by god I want it to be research worth doing! (taps the desk emphatically with each point) Creative, noble, aspirational science!

JEFF
(timidly, after a moment of silence)

Look, I'm not sure that foundations research is more creative than other fields. I agree this whole setup is unfair, but so is life. I am sure about one thing that I think we can all agree on. Foundations research is unpredictable in a different way than quantum technologies are. Yes, we could change the world's perspective on quantum science in a year. Or we could not make any significant leap in twenty years. Can we agree that foundational research is not a good fit for a five-year timeline and one chance at a future Earth?

(brief pause)

STEF

Seems reasonable——

                        ELISABETH
               (interjecting, aside)
Unless someone has a peer review manual for the end of the world committees we can check.

                          STEF
——do we have a majority?

    (Miguel nods quickly. Jeff raises his hand, then brings it down.)

                          JEFF
I thought we needed a unanimous vote.

                          STEF
Any votes for foundations? Schmidt?

    (Silence, as Schmidt raises his hands, shakes his head.)

                          STEF
No? Well then, one more proposal down. (stands up) Coffee?

                       END OF SCENE

<u>Scene 5: First coffee break</u>

AT RISE:                  Puck enters pushing a coffee cart, dressed up as a barista, looking slightly disheveled from the chaos in the outside world. Schmidt stands up, picks up chalk and continues writing on the blackboard where he left off, steadily rising in excitement.

                        STEF
        (grabs coffee absentmindedly)
Two proposals down. I can't wait to get to the good stuff. We'll seal the deal soon, I'm sure.

                    ELISABETH
        (while grabbing her own coffee)
Seal the deal?

(Stef and Elisabeth walk back towards the table. Miguel and Jeff step up to the cart and immediately recognize Puck, laughing and starting a conversation in the background.)

                        STEF
        (smug)
Yeah, seal the deal on saving humanity.

                    ELISABETH
        (confused)
I think that will still take a while; committee meetings usually do.

                        STEF
Oh, right. It's been quite a while since I sat in one of these. Board meetings are quicker. Efficient. Time is money and all that.

                    ELISABETH
Money won't matter much if we don't get this right. So we'll take our time.

                        STEF
We'll see about that. What is Schmidt doing?

>                    (Stef and Elisabeth turn to look at Schmidt, who is now
>              frantically writing on the blackboard, then sit down and start
>              looking through documents on the table. Concurrently, Jeff and
>                               Miguel laugh loudly with Puck.)

                                  MIGUEL

So you're telling me that you've been doing this at Stef's workplace every week and she still hasn't noticed?

                                  PUCK
>                    (obviously pleased with themselves)

I don't think Stef even knows that there is a barista. She thinks coffee just appears in her hand by magic. Elisabeth is the same, though for completely different reasons. All concentration, no spatial awareness.

                                  MIGUEL
>                         (courteously)

But Jeff noticed.

                                  JEFF
>                       (blushing slightly)

I think you said it first! (to Puck) Would you mind telling us a bit more about what's going on now that you're here?

                                  PUCK
>                          (nicely)

No. I'm in quite the rush. And you are all supposed to be in a rush too. (Finishes an espresso and hands it to Jeff.) But I just couldn't resist checking to see if everything is okay.

                                  MIGUEL

Sure, just one quick thing. (speaking very quickly) Do you know anything else about all of this? What do you think the aliens would prefer? Do we even stand an actual chance? Why quantum?

                                  PUCK

Miguel, your concerns are natural. They are shared by politicians, citizens, and, above all, other scientists. The world outside is in chaos. The cosmologists are rioting.

> (Cosmologists' riot noise and slogans are heard from backstage.)

                                  PUCK

>             (putting a hand on Jeff's and Miguel's shoulders)
> I must hurry away! I have full faith in you. (handing out a second coffee) Now, here's an espresso for you, Miguel. Hope you like the taste of coffee for once.

> (Schmidt goes back to check the stack of proposals, then returns to the blackboard, reaching the end of the calculation.)

>                    JEFF
>             (to Miguel, confused)
> You don't drink coffee?

> (Puck uses this moment to vanish, leaving the coffee cart behind.)

>                    MIGUEL
>             (dismayed)
> And they vanish without answering a single question. (to Jeff, shyly) I… you know, actually, I usually drink flat whites. On ice.

>                    JEFF
>             (flirtatious)
> You'll have to come by my office at some point! Just last week we got a top-grade espresso machine. Then the aliens, like thirty-five missed calls from Puck, and here we are, discarding proposals like I down coffees.

>                    MIGUEL
> So you're really passionate about coffee, huh? Can't say I'm surprised. But I'd love to learn how you prepare it!

> (They walk back towards the table.)

>                    JEFF
> Oh yeah, definitely. I'd be more than happy to, but honestly, the secret is just good beans.

>                    MIGUEL
>             (curious, but also clearly flirting)
> Where do you get them from?

>                    JEFF

It's actually Puck who sends me beans every so often. I do wonder where they're from…

                        SCHMIDT
            (turns away from the blackboard and proclaims
            loudly)

I've got it!

                    END OF SCENE

## Scene 6: Quantum simulators

(Everyone turns to look at Schmidt.)

                  ELISABETH
What do you mean, you've got it?

                  SCHMIDT
We should just build large-scale quantum computers with minuscule error rates.

                  MIGUEL
What do you mean "just"?

                  STEF
        (trying to decipher the board, which the other
        council members don't yet pay attention to)
That estimate looks right to me.

                  JEFF
We haven't even read that proposal yet! And it's not even top of the list. (Reading the stack of proposals on the desk.) Quantum simulators, (flip) quantum sensors, (flip) then finally large-scale, universal, error-corrected quantum computers.

                  STEF
It's about time we had a strong proposal.

                  ELISABETH
If you want strong proposals, then we should be talking about simulators or sensors!

                  MIGUEL
We have to talk about them all eventually. Does it really matter, the order? It's more important that we just start!

                  JEFF
Then let's go with the order Puck recommended.

                  ELISABETH
        (quickly picks up the next proposal)
Why did scientists first propose the idea of quantum computers? To optimize finance? No. To accelerate machine learning? No. Their dream was to build experiments to learn more about quantum

systems. This is where our focus should lie: Using quantum to study quantum.

We don't need quantum computers that can do everything. It is enough to have devices that are really good at one type of problem: **Quantum simulators.** (speeds up, excited) Quantum simulators are quantum systems specially designed to mimic more complicated quantum systems we want to study. They are like the early task-specific classical computers, but they are also real quantum systems we can do experiments on. Some of these experiments will help us understand exotic phases of matter. Imagine unraveling the mysteries of room temperature superconductors! Some simulators will act as quantum computers for specific problems, such as solving models that shed light on fundamental forces of nature.

Thus, simulators will have an impact in physics, in chemistry, in materials science, and in biology! In five years, we can be realistic while still dreaming big. Let us use quantum to simulate quantum!

(End of proposal.)

JEFF
Hang on, is this proposal talking about early noisy quantum computers? Or condensed matter experiments? Or something completely different?

ELISABETH
It's all and also none of the above. (excited, stands up) Quantum simulators are truly a new instrument for scientific inquiry.

MIGUEL
Here we go…

ELISABETH
I always carry these around for situations like this. (Looks through her backpack, eventually pulling out an empty egg carton and some marbles.) Okay, for one example of a quantum simulator, picture something like this.

JEFF
(amused)
An egg carton?

ELISABETH
A surface with periodic dips, made by intertwining powerful, finely tuned lasers. Now, individual atoms — represented by these marbles — can be trapped in each of these dips. (Picks up marbles and places them in the grooves.)

MIGUEL
It's called an optical lattice. Won a Nobel Prize!

(Schmidt has fallen asleep. Stef walks to the blackboard to scrutinize Schmidt's calculation. Jeff is still attentive.)

ELISABETH
The point is that even though this might look pretty simple, at these tiny scales you have to take quantum effects into account. The atoms talk to each other, and we can model the properties of the lattice — I mean, the egg carton (fiddles with the egg carton, trying to stretch and bend it) — in order to capture the details of a quantum system. Like a molecule or a protein, for example.

JEFF
So, the egg carton is a quantum simulator? But not a full-blown quantum computer?

ELISABETH
Exactly! Simulators still contain a lot of the noise that we were talking about earlier in the NISQ proposal, but the hope is that it's less of a problem with simulators. After all, the point is to simulate and better understand quantum phenomena that happen in nature. And in nature, there is noise! So it's okay for the simulator to be imperfect.

JEFF
But eventually, this egg carton could become a quantum computer?

MIGUEL
If we keep making progress on the error correction and all that.

ELISABETH
True. But that's half of why these simulators are extremely exciting! Of course, we all have colleagues who think large fault-tolerant quantum computers are decades away or impossible.

But in a world where quantum computers never work, quantum simulators still might! Just look, we are controlling all these atoms and can finally understand how they interact!

          JEFF

I see how we could learn stuff about quantum chemistry or quantum many body physics. But that's not really going to be enough to impress these aliens, is it?

         ELISABETH

You're not seeing the whole picture! We might figure out why some materials conduct electricity with zero resistance at room temperature! Think of how that would change the world!

          JEFF
      (with intrigued surprise)

Oh…?

         SCHMIDT
      (waking up)

Are they really easier to build than full-blown, error-corrected quantum computers?

  (Uncertain pause. The council members look between Stef and Elisabeth.)

          STEF

Easy is not the goal here. (vigorously) You're all being small-minded. We have the *whole world's support*. We can do literally anything! Why confine our ambition? If we try something grand and we fail, it sucks. But if we don't dream big enough, we are failing from the beginning!

         ELISABETH

So we should pick the proposal that's profitable to companies rather than academic labs, right? Companies like yours?

          STEF

That's not what this is about.

         MIGUEL

It sure does seem like that's where you want to end up.

         ELISABETH

>                    (pointedly at Stef)
> Or has this panel meeting gone on too long for your schedule?
>
>                         JEFF
> Hey, sounds like we can't make a decision about this yet. Let's just move on to the other proposals.
>
>                         STEF
> That's a waste of time when Prof. Schmidt has already identified the solution.
>
>                         MIGUEL
> Wait. How did you come to decide on quantum computers, Prof. Schmidt?
>
>            (As they all look at the blackboard, Elisabeth comes to a
>                               realization.)
>
>                         ELISABETH
> Tell me you didn't.
>
>                         SCHMIDT
>                    (excited)
> I got in early, so I analyzed all the proposals, estimating their probability of success within five years, conditioned on receiving unlimited economic and political support. It's just a simple exercise, approximating every technical hurdle I could think of as a random process, like flipping a coin, but accounting for some extra things. (to Elisabeth) Including the social impact and repercussions, of course.
>
>                         ELISABETH
>                    (with sarcasm)
> Oh well, that's alright then. Just as long as we aren't staking the Earth's future on a simplistic mathematical approximation *that doesn't account for social impacts.*
>
>                         STEF
> Look, never mind the ethics. The ambition of this project more than compensates for its technical challenges. You've confirmed my gut feeling, Prof. Schmidt.
>
>                         JEFF
>                    (jokingly to Stef)

If building quantum computers were as simple as a few lines on a chalkboard, wouldn't you have hired someone to do that years ago?

                    SCHMIDT
Well, yes, certainly *building them* might involve some engineering challenges. But predicting the likelihood of them being built, given the scale of the resources the grant recipients will have at their disposal, well that's much simpler.

                    ELISABETH
Simple?!

                    MIGUEL
Let's hear the proposal before we take this any further, shall we?

                    JEFF
But sensors were supposed to be first. (glancing around) Oh alright, here is the quantum computing one.


                        END OF SCENE

### Scene 7: Quantum computers

(Miguel reaches for the proposal, but Stef beats him to it.)

                    STEF
The holy grail of quantum computing — it is within reach. What will the world be like once humanity has built a
            (bold-faced words in this scene are accompanied
            by an imposing sound effect)
**large-scale, universal error-corrected quantum computer**?

It will be changed forever.

                    SCHMIDT
            (taking over, reading intently)
Remember the I and S in those NISQ devices? Intermediate-scale? Let's aim beyond that. **Large-scale** computers are what we want. Millions of atoms or photons encoding hundreds of thousands of packets of information; that's our goal.


                    ELISABETH
**Universal**! Some will say that finding more algorithms for problem specific devices is all we can do. Instead, let us push for quantum computers that can be programmed and reprogrammed, tackling many important problems with just one device. Seeking to change the world with quantum algorithms is futile without ensuring we have high-quality quantum computing hardware!

                    MIGUEL
            (slightly skeptical)
And finally **error corrected**! Mistakes happen to everyone, even to quantum computers. But there are ways to correct its faults! The further we lower the error rates of quantum devices, the longer our quantum computer can run for, and the more complex problems we can solve!

                    JEFF
These computers could open new horizons in chemistry and the pharmaceutical industry, help us develop new materials with useful properties, and find the optimal routes and schedules across a supply system. However, before the technology can become a household name, there are major feats of engineering to

accomplish. We did it for today's supercomputers, and we'll do it again for quantum computing!

                      STEF

The race to know which hardware design will be best is still open. An atom, an ion, a photon, a superconducting circuit? They could all be the building blocks of our future… (with full enthusiasm) **large-scale, universal, error-corrected quantum computer**!

                 (End of proposal.)

                    SCHMIDT
             (matter of fact)

As you can see, building these quantum computers will simply do more than any of the other proposals!

                      STEF

It will be a new industrial revolution.

                      MIGUEL

And a scientific one, too.

                    ELISABETH

I agree there is huge potential. *If* it works.

                      JEFF

How big of an "if" do you think it is? Some of my colleagues in computing think that it's almost inevitable now.

                      STEF

Given the breakthroughs in error correction we've seen over the past three years, I'd say we're on track to make it happen.

                    SCHMIDT

The theory is already in place! We know how to encode information in a larger codespace and only the implementation is left to work out.

                      JEFF

Do I have it right, that we have to sort of "hide" information about one particle in several particles?

                    MIGUEL

Yes! We might need tens or hundreds or thousands of physical particles for one "unit" of information to be reliable on a quantum computer.

                    JEFF
Doesn't that make the number of particles you need to use enormous?

                    MIGUEL
Yes, and don't forget about the post-processing costs. Quantum error correction needs to solve an algorithm on a classical computer. And, that algorithm takes a long time to run… and a lot of energy.

                    STEF
Sure, it's a massive cost now, but it won't look so intimidating in five years!

                    JEFF
Okay, then let's be very optimistic for a moment. If we manage to build these fantastical computers, what do we truly know we could do with them, apart from using Shor's algorithm to break RSA cryptography? I'd like to hear each other's takes.

                    SCHMIDT
We'd figure out how to use them to defend ourselves against the aliens!

                    ELISABETH
          (concerned)
And once we've dealt with the alien threat, we'd break national security across the global south? Steal everyone's credit card details?

                    STEF
          (with a wave of the hand)
Now you're just being dramatic. We have known how to design quantum-safe cryptography for years. We already use them!

                    ELISABETH
          (cynically)
You have them over in your big tech corner. Loads of smaller companies cannot afford to protect their data that well. It just extends the divide between the rich countries and the global

south, where not even governments can keep up with new cryptography protections.

                    STEF
               (confidently)
You would prefer us to have done what? Put our feet up and wait for China to intercept and store our data until they can break RSA? (pauses) And anyways, these methods are open. Everyone is free to implement the new standards.

                    ELISABETH
               (hotly)
As if everyone can afford to hire people with sufficient expertise.

                    MIGUEL
Or maybe we can move on to a, you know, less controversial example. Think of chemistry. These quantum computers could increase the resolution of calculations to what every electron is doing in a chemical reaction! If you use a classical computer to solve a chemistry problem at that resolution, you're pretty much stuck at twenty electrons. Twenty! With a large-scale quantum computer, we could finally go beyond that and understand in detail what happens when atoms bond or when molecules interact, like, you know, how different medicines interact with tissue in our bodies.

                    STEF
Well put, Miguel. I couldn't agree more.

                    ELISABETH
We could get better medicines, or harsher pesticides, depending on who is using the computers.

                    STEF
               (curt)
If you're such a technology cynic, Elisabeth, why are you even here?

                    SCHMIDT
Other than running one of the most advanced trapped ion labs in North America?

                    ELISABETH

>                    (pause, delivered with passion)
> If we were to vote for this proposal, and it does work out, then it's an opportunity to learn from the mistakes of former technologies! It would be a global effort, and we'd need to ensure socially valuable applications get priority.
>
>                         STEF
>                    (still dismissive)
> Mistakes? What mistakes? (to Elisabeth) I'd like to hear what's so wrong about the backbone of modern societies.
>
>                         MIGUEL
> Really? You don't see any mistakes? The climate emergency? Low-paid resource extraction jobs? The solidification of an imperialist-capitalist-patriarchal hegemony through unequal technological domination?
>
>                    (A pause.)
>
>                         JEFF
> That seems a bit extreme——
>
>                         SCHMIDT
>                    (interrupting Jeff)
> Why are we talking politics? The point is very simple: quantum computers are the best technology that we can imagine. And now we have a chance to actually make it happen.
>
>                         ELISABETH
> I think Miguel meant to say we need safeguards to ensure that quantum computing does the most social good for all the societies supporting this work and not just for those rich enough to build a quantum computer once they're designed.
>
>                         STEF
>                    (on the edge of her seat)
> Get back to reality - we're in a phase of disruptive innovation! Imposing regulations would kill the future before it's born. People need to be creative; people need to be free!
>
>                         ELISABETH
> I prefer to see regulations imposed before people get harmed.
>
>                         STEF

I'd prefer to get a strong project finished before the Earth gets literally destroyed.

(A moment of silence.)

MIGUEL
I definitely wouldn't want, you know, the military using my ideas.

SCHMIDT
(to Miguel)
Would you still say that if your country were under attack? I'd want the best defense possible.

STEF
We're all literally under attack right now by these aliens! And I do want the best defense possible. And whether you like it or not, if a universal quantum computer is going to be built, it will be by a group of companies working together, not in some outdated lab in the basement of an old university! Just look at the pharma industry. The medicines that we have wouldn't have been developed by academics. Ever!

ELISABETH
The *pharmaceutical industry* is your example of industry ethics?

MIGUEL
Aren't we getting a bit far from the point? The question was whether we can build these large-scale, universal, low-error quantum computers quickly. Isn't it less risky to concentrate on something more concrete, like designing a quantum computer that can do one task really well?

SCHMIDT
No, look, it's on the board! Quantum computers for specific simulators are essentially made of the same stuff as the ones made for general calculations. Why would anyone waste days on a small lemma when they could instead prove the theorem and have the lemma as a consequence?

MIGUEL
I'm lost here…

STEF

Meaning, why design a machine to do one task really well if you could aim for a more complex machine that does a hundred different tasks equally well.

                           MIGUEL
                  (somewhat exasperated)
I get that, but it's much harder to build a general-purpose quantum computer than it is to build a task-specific simulator! The question is whether progress in quantum error correction will succeed as well as we hope on a very short timeline. (shakes his head) In all honesty, staking the world on it in only five years seems extremely risky.

                           STEF

Engineering problems can be solved with enough dedication and genius, and we have the money to command both.

                         ELISABETH

Well, that's part of the problem with this proposal, isn't it? This project would rely on companies like yours, would evolve on corporate timescales, would give patents to industry partners…

                           STEF

Why is that a problem? We're at the top of the industry.

                         ELISABETH
                    (with resignation)
That's true, you might be. Only a handful of academic labs have the staff and funding to build quantum computers. And even academic labs end up outsourcing some of their equipment, when their staff aren't getting poached by companies like yours, Stef!

                           STEF

Good for them. Would you prefer they go work in consulting? I offer them money *and* fun physics.

                         ELISABETH

So, we all take on a huge risk and only large companies like yours stand to benefit if the Earth survives?

                         SCHMIDT

I'm beginning to think there was a flaw in my calculations.

                         ELISABETH

Welcome to the club.

                        JEFF

Hey, it's not a deadlock yet. We still have sensors left!

                      MIGUEL

True. I guess there's no point in voting right now. We're too far from a consensus, and we're all a bit too worked up. Let's just take a break. I'm sure some coffee will help.


                   END OF SCENE

## Scene 8: Second coffee break

(Elisabeth pulls out an agenda and starts scribbling. Jeff and Miguel stand up and chat on the side. Stef walks towards Schmidt, who is at the blackboard, carrying two cups of coffee.)

STEF
How are the calculations going? Need some brewed concentration? (offers a coffee to Schmidt)

SCHMIDT
I'm almost done, but thanks anyhow… (to himself) There's so much to revise in this calculation. I don't know what I was thinking…

STEF
I just wanted to say, even though I wouldn't stake the planet on unifying gravity and quantum in five years, I sure hope that foundations research has a place in whatever we choose. I think people have this idea that if you want to do pure science, then you have to work on quantum simulators or foundations. But really, we need people like you in quantum computing. People who think about what exactly quantum is.

SCHMIDT
I'm listening, but I'm skeptical you actually know anything about foundations.

STEF
Nonlocality, contextuality, magic… Working out what makes quantum different at its heart… We care about the same questions, really, you and I. The ingredients that make quantum truly different; they are the secret sauce that will fuel our quantum revolution!

(Elisabeth looks up from her writing and listens intently)

STEF
Isn't it beautiful? This marriage of the deep and the practical? We can work together to make quantum computers the best they can be, even if quantum gravity takes longer. We can turn paradoxes into technology. That, my friend, is what will astonish the aliens. We can blend hardware and foundations, although of course we would be changing the proposals a bit to do that. Might need a visionary thinker to pull it off.

>                (Miguel and Jeff return towards the table and blackboard,
>                                     listening.)

                              STEF
> Now that I think of it, you would be just the man for the job!
> I'd be happy to recommend you, and with the two strongest
> members of the "save the world" council, plus my company
> throwing in some extra resources, I'm sure we could manage it.

                              MIGUEL
>                (wandering into earshot and speaking through a
>                croissant to Schmidt)
> She's trying to bribe you.

                              SCHMIDT
>                (shakes off his emotional moment, loudly addresses
>                Stef)
> You are trying to bribe me! You thought you could play upon my
> enthusiasm to get my support?

                              JEFF
> Man, how are we ever going to reach a consensus?

>   (Stef walks past Elisabeth to take a seat. Elisabeth gives Stef
>                           a disgusted look.)

                              STEF
>                (irritated)
> What are you shaking your head at me for, Betty?

                              ELISABETH
> We are very different people.

                              STEF
> Well congratulations to you for stating the obvious.

                              ELISABETH
> To keep stating the obvious, you didn't quite seal your deal
> there, did you?

                              STEF
>                (angry pause)
> You just wait and see.

END OF SCENE

## Scene 9: Quantum sensors

(Jeff, Miguel, and Schmidt settle back down at the table.)

MIGUEL
(pretending there was no commotion)
Well, Prof. Schmidt? Did you reach a conclusion?

SCHMIDT
(still a bit huffed up)
Yes! It's a toss-up between quantum sensors and quantum simulators. I will need to redo the calculations with another order of precision.

JEFF
Oh, finally!

STEF
Sensors? You can't be seriously suggesting we spend a world's worth of funding on measuring the Earth's magnetic field?

SCHMIDT
Actually, I believe I'm suggesting we spend a world's worth of funding on *science*. We already use quantum sensors in gravitational wave detectors. Imagine how many fields of science could be pushed forward with better measuring equipment!

ELISABETH
That would be great for medical physics!

STEF
This is ridiculous. We already have quantum sensors! I could buy you a SQUID right now!

SCHMIDT
A squid?

MIGUEL
(jokingly)
I wouldn't mind some seafood for dinner. I'm starting to get hungry.

STEF
(annoyed)

No, not octopi. I mean **s**uperconducting **qu**antum **i**nterference **d**evices; they've been around since the 60s!

                    JEFF
               (timidly)
Sure, SQUIDs are great! But quantum sensors can mean much more than that. There are plenty of industry applications where adding more sensitive sensors would have a big impact. There are already some promising quantum prototypes in university labs. Ones that actually involve entanglement, for example.

                    STEF
               (irritable)
So tell me then, in your expert opinion, what kind of sensors would you like us to fund? Diamond-vacancy centers, squeezed light, quantum illumination?

                    ELISABETH
               (equally irritated, interrupting Jeff's move to
               speak)
That's not his job. We're here to judge proposals, not to make them.

                    STEF
But since he's so enlivened by this topic, I'd like to hear why.

                    ELISABETH
               (sarcastically, picking up the sensors proposal,
               interrupting Jeff again)
Your wish is my command, Stef. The natural sciences are first of all **empirical sciences**. Every century of science brings better instruments with which to push the precision of our inquiry. New measurements give rise to new insights — the next frontier lies always in the next decimal digit. New equipment means new eyes, and quantum sensors are the next step in this honorable tradition. (Passes the proposal to Miguel, who picks it up happily.)

                    MIGUEL
There are still theoretical and engineering challenges involved in building quantum sensors, and even more in getting them out of the lab. But they can already be used in specialized scientific experiments. As a bonus, quantum sensors can be used to improve the hardware designs of quantum computers!

> JEFF
> *(takes over the proposal from Miguel and reads with excitement)*

We know the limits of sensors that don't use any quantum effects, and we know that sensors which use quantum-specific effects like entanglement can do better! Sure, the tasks sound basic – more accurate detection of light, more precise measurements of magnetic, electric, and gravitational fields — but with these basic ingredients, we can have more precise clocks, more detailed scientific experiments, and maybe even sharper sensors for use in the mining industry and in medical technology. This field is already on its way to winning a Nobel Prize. If we push it further, impressing the aliens is guaranteed!

> *(End of proposal.)*

> STEF

Prof. Schmidt, don't you think that was a bit underwhelming? I'm sure you'd prefer something with a bit more theoretical substance.

> ELISABETH

Come on. Scientific progress isn't always about radical new ideas. Just as often it's about incremental improvements to equipment. (with resignation) Unexciting as that may be.

> MIGUEL
> *(looking to Schmidt, who was going to speak but instead nods along in agreement)*

But quantum sensors involve a lot of theory. You know, bounds on measurement precision, guiding experimental designs, measuring open quantum systems – that should be enough to keep even Prof. Schmidt happy.

> JEFF

Sure, but even if that weren't true, why does everything have to be so focused on theories with you all?

> STEF

Because they think it's more fun.

> JEFF

But without the technology to test theories, you might as well be stumbling around an infinite swamp. And without weird experimental results in a lab, we wouldn't get new theories in the first place. Seriously, we only have quantum science because people bothered doing experiments and doing them more and more precisely!

                    ELISABETH
Well, you've got me there. But I still think simulators are more interesting.

                    JEFF
Does anyone have concerns that don't reduce to "I think sensors are boring?"

                    MIGUEL
I have concerns about some of the commercial and military applications. It's not like these sensors will allow anyone to see through walls, but I mean using sensors in mining and oil extraction? That's like saving the Earth in five years to boil it in ten.

                    ELISABETH
             (enthusiastically)
Solid poetic injustice though.

                    JEFF
Poetic injustice? Do you even hear yourself right now?

                    ELISABETH
Yeah and I can see myself too. All of us, stuck in a room, bickering over the fate of the world through research proposals. Excuse me if I prefer comedic deflection to absolute panic.

                    SCHMIDT
Can we just get back to the science? The part we're all *actually* qualified to do?

                    MIGUEL
Okay, sure. This is my take: there are so many areas of science where pushing for just a little more resolution, a little more precision, could make a huge difference.

                    JEFF

And not just in specialized lab environments! Companies like mine are starting to market advanced sensors that work outside the lab. We've taken silicon photonics research and commercialized it towards personalized medicine. It should be the same with the next generation of quantum sensors.

MIGUEL
Maybe eventually? But for me, the beauty of this proposal is that some scientific applications work even if we can't get the hardware to the stage of commercialization within five years, you know? Getting scientists better interferometers could lead to huge advancements.

ELISABETH
And we can add a clause to the proposal about excluding military use cases.

STEF
(sarcastically)
Right, because scientists are paragons of virtue.

JEFF
And moving on, I'm still confident those scientific sensors will help us move towards more stable, smaller, easier to operate designs. And those can lead to specific commercial applications.

STEF
So early quantum hardware can't be of any use but early sensors can be? I liked it better when you were all moony- eyed over quantum simulators.

JEFF
We're not saying quantum computers don't have some early applications. *I'm* just thinking, with the whole world on the line issue, it feels better to play it safe. (looking to Miguel for support) Sensors have a balance of applications that will probably work and some that are more ambitious.

STEF
You don't know that! You're all acting like sensors and simulators are so much easier to design than error-corrected quantum computers! Well, I've worked on quantum hardware for twenty years and you know what? We have the same problems, sometimes even the same atoms! Just because big quantum

computers have "big scary business" applications you don't like them?

                        ELISABETH
*Yeah*. Science funding should go to scientific applications first. Look, we can be ambitious with simulators or delusional with computers.

                        STEF
Oh, spare us your self-righteous cynicism.

                        ELISABETH
Cynicism? For error correction on large quantum computers to work, everything has to go right, as optimistically as we can imagine, for five years of experimental work. Smaller, specialized simulators can handle more problems along the way.

                        STEF
That's simplistic. We're all rolling dice here whatever we do, and we need to impress the new cosmic order. That's what we should pick the proposal based on.

                        JEFF
Hey, no one is saying no to the big quantum computers yet.

                        ELISABETH
Oh, I am, now that I see the kind of person who supports them.

                        MIGUEL
              (quietly)
So helpful.

                        ELISABETH
*And* they agree with me, (gestures at Schmidt, Jeff) so you're outvoted.

                        JEFF
              (to Stef)
I just don't think it's the safest option we have right now.

                        STEF
You can't all seriously think I'm going to stand by and let a bunch of naïve academics end the world because you want to

pretend you spent your life on some idealistic search for pure knowledge.

                    MIGUEL
Let's all take a step back here.

                    ELISABETH
No, I'd like to hear her out. Please explain how out of touch and naïve I am for wanting to use public funding for the public instead of your company.

                    STEF
You think you have a job, a lab, because the whole world is dewy-eyed about the exciting world of photons and wants to send little notes of encouragement to each other through the quantum internet? No, you have DARPA funding because they want to have precise timing on the battlefield and send encrypted information during wars. Companies and militaries and states: that's who pays for your pure discovery. That's how it's always been, right back to the days of your beloved Bohr and Oppenheimer.

                    MIGUEL
             (with great dismay)
DARPA? You work with the U.S. Military?

                    ELISABETH
My *student* has DARPA funding. She'd have to leave the country without it. The world's not a black and white place like your type likes to think.

                    MIGUEL
My type? What is my type?

                    ELISABETH
Young? Idealistic? Full of useless theorizing and an inability to compromise.

                    MIGUEL
You didn't seem to think it was so useless when I agreed with you. And *you're* coming at *me* about being uncompromising?

                    ELISABETH
I——

MIGUEL
You've spent half this session poking at industry, and it's not just righteous anger, is it? You're ashamed of all your own compromises along the way.

ELISABETH
You can't understand when you don't have students looking to you for funding, for security.

MIGUEL
*I have students*. I have a lab, I have grants, I have committee roles, the whole thing. What I don't have is the exact same ideology as you, and maybe I push more rules than you do. That's because people like you paved the way for me to get away with it. So, respectfully, why don't you tell me what I should do so that you treat me as an equal?

SCHMIDT
Come now, this is getting ridiculous!

MIGUEL
Complete deference? A signed recommendation letter from Marie Curie? What?

ELISABETH
Actually, I think that rant might have done the trick.

| MIGUEL | SCHMIDT |
|---|---|
| What? | What? |

ELISABETH
That. Standing up for yourself. Apparently we all need some sense yelled into us from time to time.

SCHMIDT
Are you kidding? You needed him to air grievances in a committee meeting?

ELISABETH
What, like we haven't both seen worse in committee meetings? (to Miguel) Keep your optimism then and keep your patience for the older generations. And even keep your vote for sensors; your scientific opinion is just as legitimate as mine.

MIGUEL

Are you serious?

                        ELISABETH

I am. I am also aging and embarrassed to find myself entrenched in old battles. Maybe sometimes a new strategy is necessary.

                        MIGUEL
           (with surprise)

That, that means a lot to me.

                        ELISABETH
           (calm)

You're still utterly wrong though; simulators are a much stronger scientific investment. And he agrees with me. (gestures at Schmidt)

                        SCHMIDT

No I don't. I prefer the quantum computing proposal.

           (Group reaction.)

                        STEF

After all that, you agree with me?

                        SCHMIDT

I almost didn't because of your bribery. And I definitely want your company banned from participating in the funding. But if we focus on science-based applications, a universal error-corrected computer would do the most across all areas of science.

                        STEF

This is the most unhinged, unprofessional decision making I have ever been part of. You all want to cast me as the villain, fine. But my company is too important to quantum technologies to be left out. Even if I'm not involved, you need our expertise.

                        JEFF

Well that depends on what we choose. We haven't made a final decision. Sure, you two seem thrilled with computers, (gestures at Schmidt, Stef, then turns to Elisabeth) and you haven't budged from simulators, have you?

                        ELISABETH

Of course not. It's not been that radical of a day.

JEFF
And I still think sensors are the best option, like Miguel, right?

MIGUEL
That's because you have good taste.

ELISABETH
I would have voted for quantum communications if that had been one of the proposals. I can start voting for that if you want more chaos.

SCHMIDT
Good point. Why was there no quantum communications proposal? Even someone my age thinks the internet could definitely use an upgrade.

STEF
(sarcastically)
I guess Betty didn't have the pockets to push communications that far.

JEFF
Stop it! This is already more chaos than we need. You say this is dysfunctional and unprofessional? Well maybe that makes sense. I mean, what is professional about five people in a room gambling on the fate of the world? Hell, this is so insane I could bet those aliens are playing with us, probably chuckling away somewhere behind the scenes.

SCHMIDT
Agreed, but who else can do it? We're the ones who know enough to make an informed decision.

STEF
More like a cowardly decision.

JEFF
We all know how you feel about pure science agendas. That doesn't help us break the tie.

MIGUEL
(wearily)

I guess we just keep talking about it? Maybe meditate, I don't know, clear our minds, and then come back to the discussion? Puck seemed in a rush though. I wonder what's going on out there…

                    ELISABETH
How? We all have different values shaping our opinions.

                    MIGUEL
So how about we just don't make a decision?

                    SCHMIDT
What?

                    MIGUEL
After all, it would ultimately be best to integrate all these technologies together, to have the data from a quantum sensor go straight into a quantum computer, for example. And we have all the funding in the world, right?

                    JEFF
But not all the time in the world, and boy does project integration need loads of that! I think we need to make a decision.

                    SCHMIDT
But perhaps we need to give up on having a unanimous one?

                    JEFF
              (picking up Puck's letter)
We can't! Puck gave us precise instructions. One. Unanimous. Decision.

                    STEF
Yeah, just after they left us a cute letter and said that we needed to pick a save the world science project. What I'm saying is, we can be a little skeptical of Puck right now.


                    END OF SCENE

<u>Scene 10: The verdict</u>

(Puck saunters in and addresses Stef.)

PUCK
Ouch? After I made you dinner last week? (addressing the group) So, how's it going?

(Elisabeth is wide-eyed with surprise, and Stef gasps.)

JEFF
I'm not sure if all the coffee breaks in the world can fix this situation.

MIGUEL
We're a bit stuck. I'm sorry. I am well aware that the clock is ticking. But we'll need more discussion time.

SCHMIDT
And possibly several pro/con charts, combined with a value-weighted ranking system.

PUCK
That well, huh? I suppose that's to be expected. But don't worry, we have all the recordings for you to look back through.

JEFF
Recordings?!

ELISABETH
Oh god.

PUCK
What, did someone not watch her temper again? If you're not careful, you might need to go back to yoga class.

JEFF
Sorry, I don't follow.

PUCK
Oh, inside joke.

JEFF

No, not that. The recordings don't help with the main problem, which is that we're still stuck.

                        PUCK
Yeah, I know. I figured you would be.

                        MIGUEL
You figured? Oh no, what are you actually up to?

                        PUCK
I needed you all to discuss the arguments candidly. Scientists are usually so careful in public, you don't want to say anything wrong. But I needed you all to express your actual opinions and have real stakes. I mean, really, why else would I put such a colliding group of personalities together?

                        MIGUEL
I'm — Well I'm probably going to be very annoyed with you later. For now, I'm relieved. But why did you need us to fight it out like this?

                        PUCK
Because Jeff was right! This is too big of a decision for just five people. I've been streaming your discussion from the beginning to the actual voters who volunteered for this. They'll vote based on their own values and the knowledge they got from you. (Touches phone, voting instructions appear for the audience.) Instructions just went out, and I can't wait to see what they all decide.

                        JEFF
We don't have to be responsible for this decision?

                        STEF
Our investors are going to see this?!

                        SCHMIDT
You don't think we can talk candidly to the public?

                        PUCK
Errr… I can't help but be a little mischievous here and there.

                        MIGUEL

Hmm, I like it. Not the blatant manipulation part, but this is too big a decision for just us.

                        STEF

This is ridiculous! *Who* is choosing? What if they choose wrong? We can't all just fly away to another planet like you can!

                      ELISABETH

I don't think it's ridiculous, maybe a bit Shakespearean though. "Oh time, thou must untangle this, not I."

                      SCHMIDT

I think you've lost all of us now.

                      MIGUEL

"It is too hard a knot for me to untie."

                      ELISABETH

*Twelfth Night*.

                      MIGUEL

Sometimes we don't have enough information to know if we're making the right decision.

                      ELISABETH

So, what's left to do except pretend that the twisted paths of history have a direction and hope that it rewards effort? That's all of life, basically.

                      MIGUEL

But it's especially true in science. You pick a topic to devote your life to, you hope it's interesting, and you hope it helps people.

                      SCHMIDT

And you accept a risk of obscurity, of failing to find the right ideas, of being the person whose work demonstrates that something probably doesn't work.

                      JEFF

Even if we do fail, I'm much happier having that burden off my shoulders. Who are we to decide what humanity should gamble on?

                      PUCK

Thou spoke it well. The Earth shall vote and researchers shall toil and onwards and upwards in time we shall run. Dear friends, let us raise the cry here first: to risk, to progress, and, just possibly, to a safer world.


END



\end{document}